# Jupiter – Friend or Foe?  IV: The influence of orbital eccentricity and inclination.

Horner, J.[1] & Jones, B. W.[2]

**ABSTRACT**

For many years, it has been assumed that Jupiter has prevented the Earth from being subject to a punishing impact regime that would have greatly hindered the development of life. Here, we present the fourth in a series of dynamical studies investigating this hypothesis. In our earlier work, we examined the effect of Jupiter's mass on the impact rate experienced by the Earth. Here, we extend that approach to consider the influence of Jupiter's orbital eccentricity and inclination on the impact rate from asteroidal bodies and short-period comets.

We first considered scenarios in which Jupiter's orbital eccentricity was somewhat higher and somewhat lower than that in our Solar System, for a variety of "Jupiter" masses. We find that Jupiter's orbital eccentricity plays a moderate role in determining the impact flux at Earth, with more eccentric orbits resulting in a noticeably higher impact rate of asteroids than is the case for more circular orbits. This is particularly pronounced at high "Jupiter" masses. For the short-period comets, the same effect is clearly apparent, albeit to a much lesser degree. The flux of short-period comets impacting the Earth is slightly higher for more eccentric Jovian orbits.

We also considered scenarios in which Jupiter's orbital inclination was greater than that in our Solar System. Increasing Jupiter's orbital inclination greatly increased the flux of asteroidal impactors upon the Earth. However, at the highest tested inclination, the disruption to the Asteroid belt was so great that the belt would be entirely depleted after an astronomically short period of time. In such a system, the impact flux from asteroid bodies would therefore be very low, after an initial period of intense bombardment. By contrast, the influence of Jovian inclination on impacts from short-period comets was very small. A slight reduction in the impact flux was noted for the moderate and high inclination scenarios considered in this work – the results for inclinations of five and twenty-five degrees were essentially identical.

[1] Department of Astrophysics and Optics, School of Physics, University of New South Wales, Sydney 2052, Australia, e-mail: j.a.horner@unsw.edu.au
[2] Astronomy Discipline, Department of Physics, Astronomy, and Space Science, The Open University, Milton Keynes, MK7 6AA, United Kingdom

# Introduction

Since the detection of the first planet orbiting a Sun-like star (51 Pegasi, Mayor & Queloz, 1995), an ever-increasing number of exoplanets have been detected. Radial velocity surveys, such as HARPS (e.g. Pepe et al., 2004, Moutou et al., 2009, Mordasini et al., 2011) and the Anglo-Australian Search for Planets (e.g. Jones et al., 2002, O'Toole et al., 2009, Tinney et al., 2011) have found the great majority of known planets, but to date, no ground based technique has been capable of finding truly Earth-like planets orbiting Sun-like stars. However, in the coming years, the *KEPLER* satellite will detect the first truly Earth-like planets around other stars (e.g. Borucki et al., 2011a, b), and the search will begin in earnest to detect signs of life upon them. In order to focus that search most efficiently, it is clearly important that the various influences on planetary habitability be considered in some depth, such that potentially habitable Earth-like planets can be compared, and searches be directed towards the most promising targets (e.g. Horner & Jones 2010a, 2011). For many years, it has been widely held that a key ingredient of planetary habitability is the presence of a distant giant planet, which could act as a "shield" for a potentially habitable world, ensuring that the world does not experience an overly punishing impact regime, and thereby facilitating the development of life upon the world (e.g. Ward & Brownlee, 2000; Greaves 2006). However, this idea had, until recently, been the subject of remarkably limited scientific study (Wetherill, 1994; Laasko et al., 2006). For a more detailed discussion of the history and origin of this widely held belief, we direct the interested reader to Horner & Jones, 2008b.

In this ongoing series of papers (Horner & Jones 2008a, 2008b, 2009, 2011; Horner, Jones & Chambers, 2010), we have examined in great detail the influence of Jupiter-like planets on the impact flux experienced by the Earth from the near-Earth asteroids, the short-period comets, and the long-period comets. In those works, we examined the way in which changes in Jupiter's mass would alter the relative rate at which objects from the three populations of potentially hazardous objects collide with the Earth. Our results were startling. For the two main populations that threaten the Earth in the current epoch (the near-Earth asteroids and the short-period comets), we found that the relationship between Jovian mass and impact rate was rather complicated. At low "Jupiter" masses, the impact rate from both populations was very low, a result of such small planets having great difficulty in emplacing objects on Earth crossing orbits. Similarly, at high "Jupiter" masses (similar to, or larger than, that of our Jupiter), the impact rate was relatively low for both populations, albeit somewhat higher than for the least massive "Jupiters". Between these two extremes, however, we found significant peaks in the impact flux upon the Earth in our simulations. For both the near-Earth asteroids (Horner & Jones, 2008b) and the short-period comets (Horner & Jones, 2009), we found that the greatest impact flux was experienced when the "Jupiter" in the simulation was between 0.2 and 0.3 times as massive as our Jupiter. In other words, our Jupiter provides only modest shielding, compared to scenarios in which it is far less massive (for example, when it is comparable in mass to Neptune), but were it around the mass of Saturn, then the impact flux at the Earth would be far greater than that we observe.

By contrast, when we look at the influence of Jupiter's mass on impact rate of the third population of potentially hazardous objects, the long-period comets (e.g. Wiegert & Tremaine, 1999, Levison, Dones & Duncan, 2001, Horner & Evans, 2002), we found that the more massive the "Jupiter", the lower the impact rate upon the Earth (Horner, Jones & Chambers, 2010). For the long-period comets, then, it seems that Jupiter really does act as a shield. However, the long-period comets are thought to contribute only a small fraction to the impacts of asteroids and comets on Earth (e.g. Chapman & Morrison, 1994, Morbidelli et al. 2002, Horner, Jones & Chambers, 2010).

*Modifying the giant planet's orbit*

But what of the orbit of the giant planet? Does that, too, play a role in determining the impact flux of potentially habitable planets in its host system? In this work, we build upon our earlier results by examining the influence of the eccentricity and inclination of Jupiter's orbit. We consider only the two main populations of impactors in this work – the near-Earth asteroids and the short-period comets. Since the long-period comets move on orbits that are essentially isotropically distributed in orbital inclination, and that are barely gravitationally bound to the Solar System, it is reasonable to assume that minor changes in the orbit of Jupiter will have little or no effect on the flux of those comets through the inner Solar System, particularly when compared to the influence of changes in its mass.

**The effects of orbital eccentricity**

In our Solar System, Jupiter moves on an orbit that is slightly elliptical (with an orbital eccentricity of 0.048775). It is interesting, therefore, to consider whether Jupiter would have more, or less, of an influence on the impact rate on Earth if its orbit was more circular, or more eccentric. We considered two different scenarios. In the first, Jupiter's orbit was significantly less eccentric than that we observe – with a value of just 0.01. In the second, Jupiter's orbit was significantly more eccentric, with a value of 0.10. It is interesting to note that, if Jupiter's orbital eccentricity is increased much beyond this 0.10 value, the outer Solar System rapidly becomes unstable[1]. As such, it was not possible to test scenarios, based on our own Solar System, where Jupiter was moving on a highly eccentric orbit.

*The asteroids*

Building upon our earlier work, we set up suites of integrations identical to those described in Horner & Jones, 2008b, in which we examined Jupiter's influence on the flux of material from the Asteroid belt to the Earth. Since the Asteroid belt and near-Earth asteroid populations we observe today are a direct result of dynamical sculpting by the giant planets over the 4.5 Gyr since they formed, it would be unreasonable to use the current belt as the basis for our test populations of asteroids for scenarios involving "Jupiters" on different orbits, or of different masses. To get around this problem, in Horner & Jones 2008b, we describe how we created, for each Jovian mass tested, a sample asteroid belt intended to represent an "unperturbed" distribution, based upon the assumption that the asteroid belt formed from a dynamically cold disk of material during the early stages of planet formation. For consistency, in this work, we performed exactly the same suite of dynamical integrations as described in that work, with just one change – rather than placing our test "Jupiters" on the orbit occupied by Jupiter in our own Solar System, we instead placed them on orbits with eccentricities of 0.01 (low eccentricity case) and 0.10 (high eccentricity case). For comparison, the current eccentricity of Jupiter's orbit as used in our earlier work lies between these two extremes, at 0.0488. We used the exact same initial distributions of asteroids, and the same range of "Jupiter" masses, in order to allow a direct comparison between our new results and those in Horner & Jones, 2008b.

---

[1] Before carrying out our detailed simulations, as described in this work, we performed a few short and simple tests to see whether the initial orbital elements we were choosing for our "Jupiters" resulted in dynamically stable scenarios. In those simulations, the orbit of Jupiter was varied, and the giant planets Saturn, Uranus and Neptune were included, moving on the orbits they currently occupy within the Solar System. Clearly, if the giant planets themselves are highly unstable, the question of the influence of those planets on the long-term habitability of terrestrial planets in the same system becomes somewhat moot, as the system itself becomes disrupted on astronomically short timescales.

Our dynamical simulations were performed using the *Hybrid* integrator in the *N*-body dynamics package *MERCURY* (Chambers, 1999). For each "Jupiter"-mass considered (namely 0.01, 0.05, 0.10, 0.15, 0.20, 0.25, 0.33, 0.50, 0.75, 1.00, 1.50 and 2.00 times the actual mass of Jupiter), we followed the dynamical evolution of our test asteroid belts, which contained 100,000 test particles, for a period of 10 Myr, under the gravitational influence of the Earth, Mars, "Jupiter", Saturn, Uranus and Neptune. As was the case in our earlier work, the Earth was once again inflated to a radius of 1 million kilometres, in order to maximise the frequency with which our Asteroids would impact upon it. Following the methodology of that work, the Earth, Mars, Saturn, Uranus and Neptune were placed on orbits identical to those they occupy in the current Solar System. The orbit of "Jupiter", too, was identical to that of Jupiter in our Solar System, aside from its modified eccentricity. For a more detailed overview of our methods, including a description of the creation of our unperturbed asteroid belts, we direct the interested reader to Horner & Jones, 2008b.

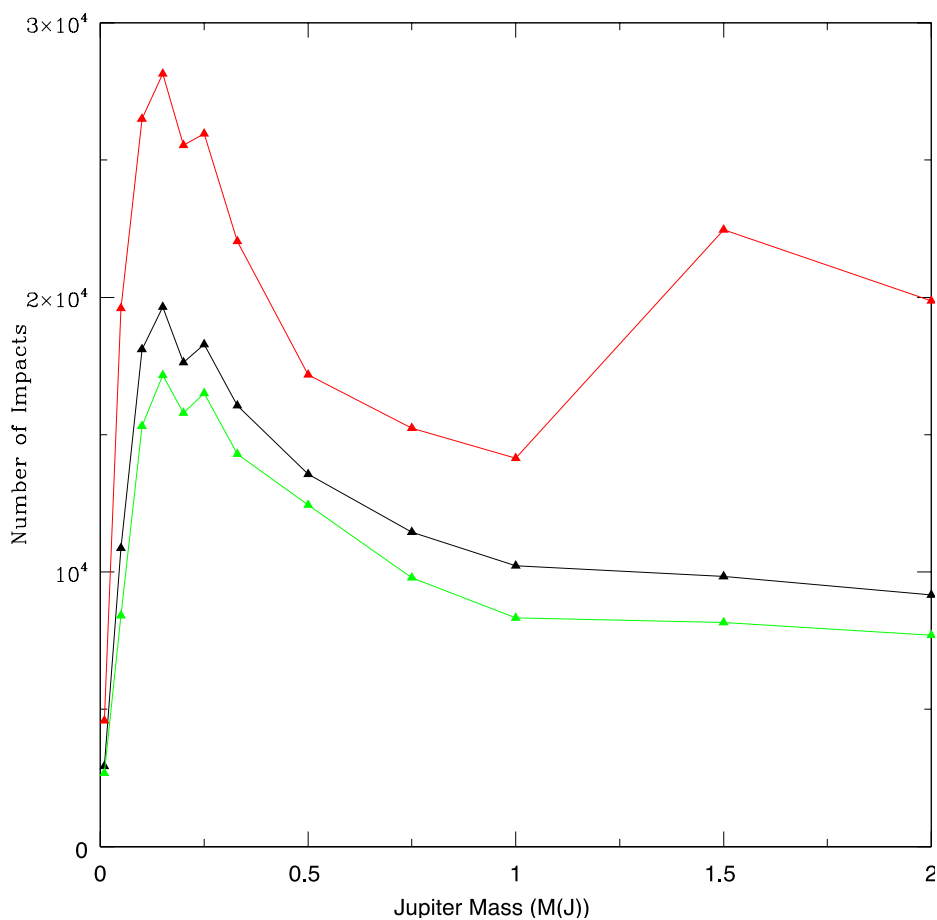

**Figure 1:** The variation in the impact flux experienced by the Earth from objects originating in initially "unperturbed" asteroid belts, as a function of Jupiter's mass. The results of our original simulations, as detailed in Horner & Jones 2008b, are shown by the black curve and black points. The scenario considered in that work placed Jupiter on its actual orbit, with orbital eccentricity 0.0488. The data plotted in green, by contrast, shows the results for simulations in which Jupiter's initial orbital eccentricity was set to 0.01 (our low eccentricity case), while the data plotted in red show the results for simulations where Jupiter's eccentricity was set at 0.10 (high eccentricity case). All other variables, and the initial distributions of asteroids, were identical between the three suites of integrations shown here.

The results of our integrations examining the influence of Jupiter's orbital eccentricity, as a function of mass, are shown in Figure 1. It is obvious that, the more eccentric Jupiter's orbit, the higher the impact rate experienced at Earth, for all "Jupiter" masses considered, with the more eccentric scenarios typically leading to an impact flux some 50% greater than the least eccentric scenarios. It is also noticeable that, at the highest masses tested (1.5 and 2.0 $M_J$), the influence of Jovian eccentricity is somewhat greater than at lower masses, with a distinct secondary peak in the impact flux appearing at 1.5 $M_J$. As can be seen in Figure 2, which shows the final asteroid belts for the three different orbital eccentricities at a "Jupiter" mass of 1.5 $M_J$, the scenario with the most eccentric "Jupiter" (plotted in red) resulted in an asteroid belt that was noticeably depleted across its entire width compared to the two lower eccentricity cases. In addition, the increased eccentricity causes the depleted region, resulting primarily from the effects of the $v_6$ secular resonance[2], to be both broader and located at a greater heliocentric distance, causing it to more heavily deplete the belt. In fact, this outward drift of that resonance was visible in the high eccentricity runs for all "Jupiter" masses, as can be seen in Appendix I, where we plot the final belts for the other "Jupiter"s tested. Finally, Figure 2 also shows that the outer edge of the asteroid belt is significantly more depleted for the scenario with the most eccentric "Jupiter" than for the others. This is the direct result of Jupiter's perihelion distance being smaller in that case, allowing it to directly perturb the belt's outer reaches. This depletion contributes to the secondary peak visible in Figure 1.

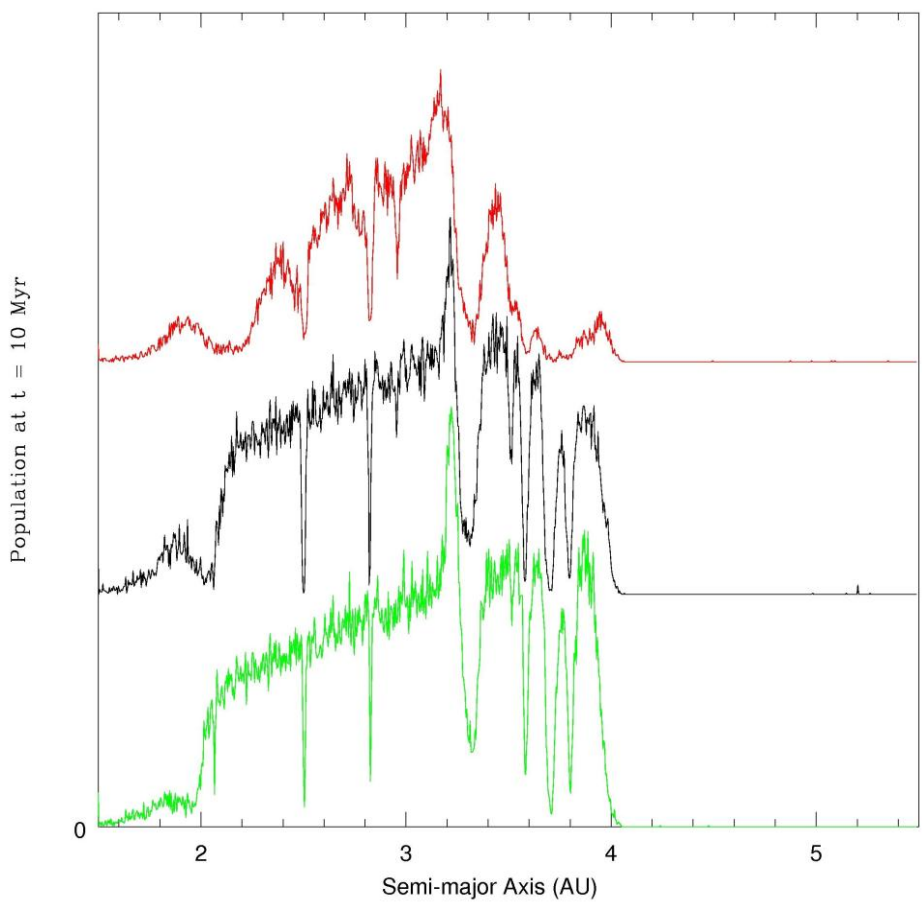

**Figure 2:** The final distribution of asteroids at the end of our 10 Myr simulations, for a "Jupiter" of mass 1.5 $M_J$. As above, the data plotted in green show results for the 0.01 eccentricity case, those in black are for the actual eccentricity (0.0488), and those in red the highest eccentricity tested, 0.10. The asteroid belt for the highest eccentricity scenario is clearly significantly more depleted at all semi-major axes than was

---

[2] In Horner & Jones (2008b), we found that the $v_6$ secular resonance was the key driver of variations in the impact rate due to asteroids as a function of "Jupiter"'s mass. As that mass increased, the location of the $v_6$ secular resonance moved inward in such a way that the maximum disruption of the asteroid belt, and hence the greatest impact flux at Earth, was observed at between 0.2 and 0.3 Jupiter masses.

the case for the other two scenarios. In addition, the influence of the $\nu_6$ secular resonance is clearly visible in the medium and high-eccentricity cases (as a region of depletion around 2 AU for the medium eccentricity case, and centred at around 2.1 AU for the high eccentricity case). It is noteworthy that the region disrupted by that resonance is located further from the Sun when "Jupiter"'s eccentricity is increased. For clarity, we note that both the black and red curves start at 0 on the y-axis, and have simply been shifted vertically in order to allow easy comparison between the three distributions.

*The short-period comets*

The short-period comets is another population whose properties, at the current epoch, are determined by our Jupiter. As such, it would be unfair to take the currently observed short-period population and use them as the basis for a study of the influence of Jupiter on the impact rate at Earth – with a different Jupiter, the family itself could doubtless look very different. Whilst the proximate source of the near-Earth asteroids, as discussed earlier, is well established as being the asteroid belt, the origin of the short-period comets is still under some debate. The proximate source of the short-period comets is the Centaur population (e.g. Duncan, Quinn & Tremaine, 1988; Horner et al., 2003; 2004a, b), a population of dynamically unstable objects that move on orbits with perihelia between the orbits of Jupiter and Neptune. The origin of the Centaurs has been widely discussed, and the population likely features contributions from a number of different reservoirs within our Solar System, including the Jovian and Neptunian Trojans (e.g. Horner & Lykawka, 2010a, b), the Scattered Disk (e.g. Duncan & Levison, 1997), the Edgeworth-Kuiper belt (e.g. Holman & Wisdom, 1993), and even the inner Oort cloud (e.g. Emel'yanenko, Asher & Bailey 2005). Rather than enter into the thorny issue of trying to model the source of the Centaurs, we instead chose to use the orbits of a number of the known Centaurs to build an "unperturbed" parent population for the short-period comets in our runs. We direct the interested reader to Horner & Jones, 2009, for more details of how we created the test population for our integrations.

In this work, we follow Horner & Jones, 2009, and perform two new suites of dynamical integrations examining the influence of Jovian eccentricity on the impact flux from the short-period comets. As for the asteroids, we used exactly the same initial conditions as those used in Horner & Jones, 2009, changing additionally the orbital eccentricity of our "Jupiter". In particular, for each "Jupiter" tested, we followed the dynamical evolution of just over 107000 test particles (based loosely on the orbits of 104 of the known Centaurs) moving initially on distant Centaur orbits. For full details of the methodology employed, we refer the interested reader to Horner & Jones, 2009.

The results of our simulations are shown in Figure 3. As was the case for the impact rate due to the asteroids, it seems that an increased orbital eccentricity for Jupiter (from its actual value of 0.0488), leads to a somewhat enhanced flux of short-period comets impacting upon the Earth. In addition, it is noticeable that the high-eccentricity scenarios shown in red feature a small secondary peak in impact flux for a giant planet of mass 1.5 $M_J$. The feature is, however, significantly less pronounced than was the case for the high-eccentricity integrations of the asteroid belt.

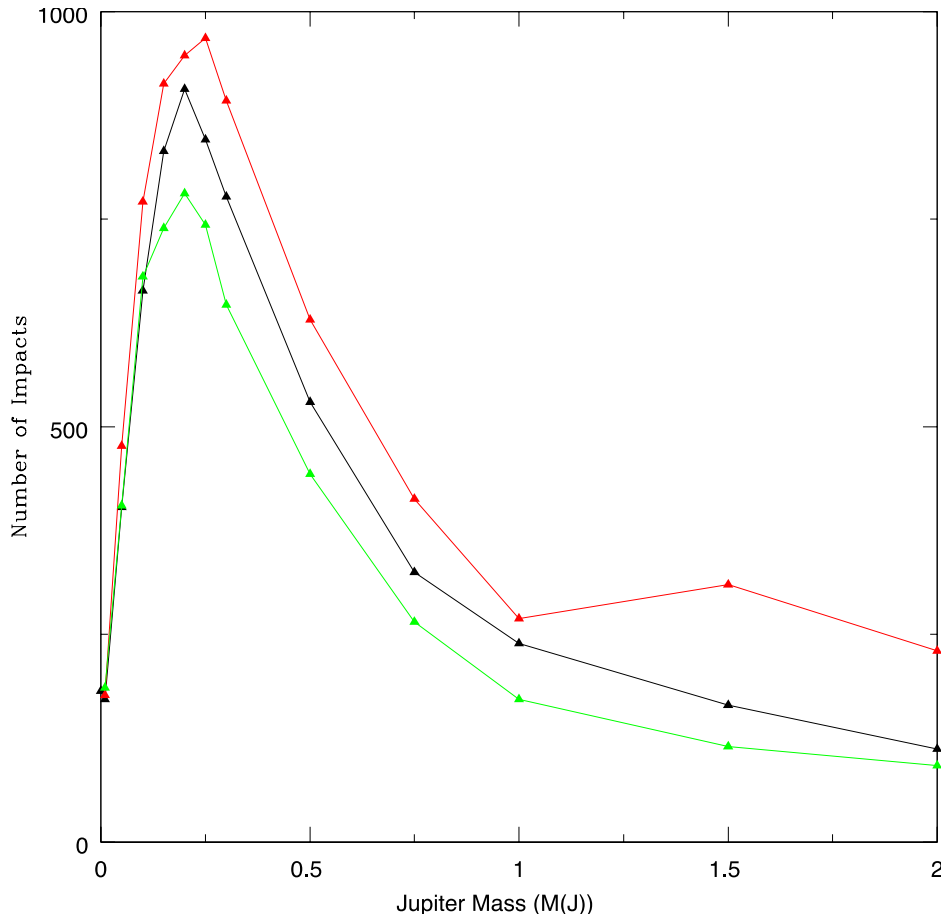

**Figure 3.** The variation in impact flux at the Earth due to the short-period comets, as a function of Jupiter's mass. As in Figure 1, the data shown in black is a reproduction of our earlier results, corresponding to Jupiter as it is today in our Solar System (Horner & Jones, 2009). The orbital eccentricity in that case is 0.0488. The green data show the results for scenarios in which Jupiter is placed on an orbit of eccentricity 0.01, and the red data show the results when Jupiter is placed on an orbit of eccentricity 0.1. As was the case for the asteroids, a greater orbital eccentricity for Jupiter results in an increased impact flux from the short-period comets, although the enhancement is perhaps somewhat smaller than that in Figure 1. It is noticeable, once again, that at larger eccentricities, a small secondary peak is visible at around 1.5 $M_J$.

**The effects of orbital inclination**

Within our Solar System, Jupiter's orbit, with an inclination of 1.31°, is only slightly inclined from the plane of the ecliptic. However, a large number of exoplanets have been discovered moving on orbits that are highly inclined to the plane of their host star's equator (e.g. McArthur et al., 2010). As such, it is clearly interesting to consider the influence of the orbital inclination of giant planets on the impact flux that would be experienced by potentially habitable worlds in the same system. In order to look into this, we again revisited our earlier work, and examined how the impact flux resulting from the near-Earth asteroids and the short-period comets would be different were Jupiter's orbital inclination greater than that we observe. We considered two separate scenarios, in addition to our earlier work – the medium inclination case, where Jupiter's initial inclination was set at 5°, and the high inclination case, where its inclination was set at 25° at the start of the integration. Aside from these modifications, we used exactly the same input parameters as in our earlier work, as described above in the section on orbital eccentricity.

*The asteroids*

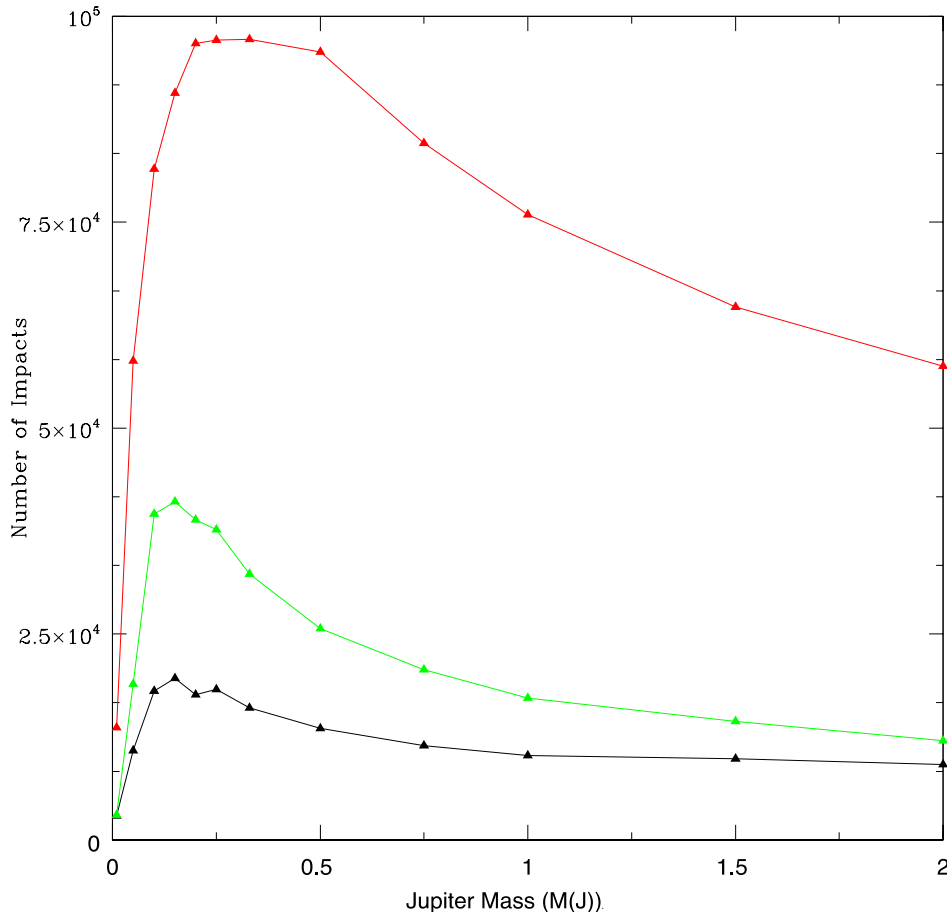

**Figure 4:** The variation in asteroidal impact flux experienced by the Earth as a function of Jovian mass. The results plotted in black were previously published in Horner & Jones, 2008b, and represent the actual case in which the giant planet's orbit is barely inclined to the plane of the ecliptic ($i = 1.31°$). The data plotted in green show the impact flux for a scenario in which Jupiter's orbit is inclined by 5° to the plane of the Solar System, while the results in red show the scenario where Jupiter's orbital inclination is set at 25°. Remarkably, in the high-inclination case, the great majority of test particles in the simulations end up colliding with the inflated Earth used in these calculations, showing that a highly-inclined Jupiter would very quickly remove any trace of the asteroid belt. For a more moderate orbital inclination, the impact rate is still noticeably increased over that detailed in our earlier work.

As can be seen from Figure 4, the orbital inclination of "Jupiter" can have a catastrophic effect on the asteroid belt. Increasing the inclination to just 5° is sufficient to significantly increase the impact flux experienced by the Earth at all "Jupiter" masses (typically by a factor of ~1.5 to 2). Increasing the inclination further, to 25°, has an even more dramatic effect, with the asteroid belts studied losing more than half their members within the 10 Myr duration of our integrations for all but the smallest "Jupiter" mass studied. Rather than suggesting that highly inclined giant planets must therefore pose an enormous threat to any potentially habitable planets within their host system, this instead suggests that the presence of a highly inclined giant planet would significantly disrupt the formation and evolution of an asteroid belt within that system to such an extent that the belt would rapidly cease to exist. Whilst this would clearly have implications for the initial accretion and hydration of potentially habitable planets in that system (e.g. Horner et al., 2009), were such planets able to form, it is clear that they would suffer essentially no impacts from putative asteroids once any belts were cleared out, potentially rendering such highly-inclined giants friends, rather than foes. On the other hand, should some mechanism excite the orbit of a "Jupiter" from being essentially co-planar with the asteroid belt to an orbit highly inclined to it, our results clearly show that this would have catastrophic consequences for any potentially habitable planets in that system at that time, since such excitation would lead to a rapid depopulation of the asteroid belt, and an associated period of remarkably intense pummelling of the planet in question. For that period, such a "Jupiter" would clearly be a foe. However, over time, the impact flux would fall to very low levels, as the belt becomes entirely depleted, eventually resulting in a remarkably clement impact regime.

*The short-period comets*

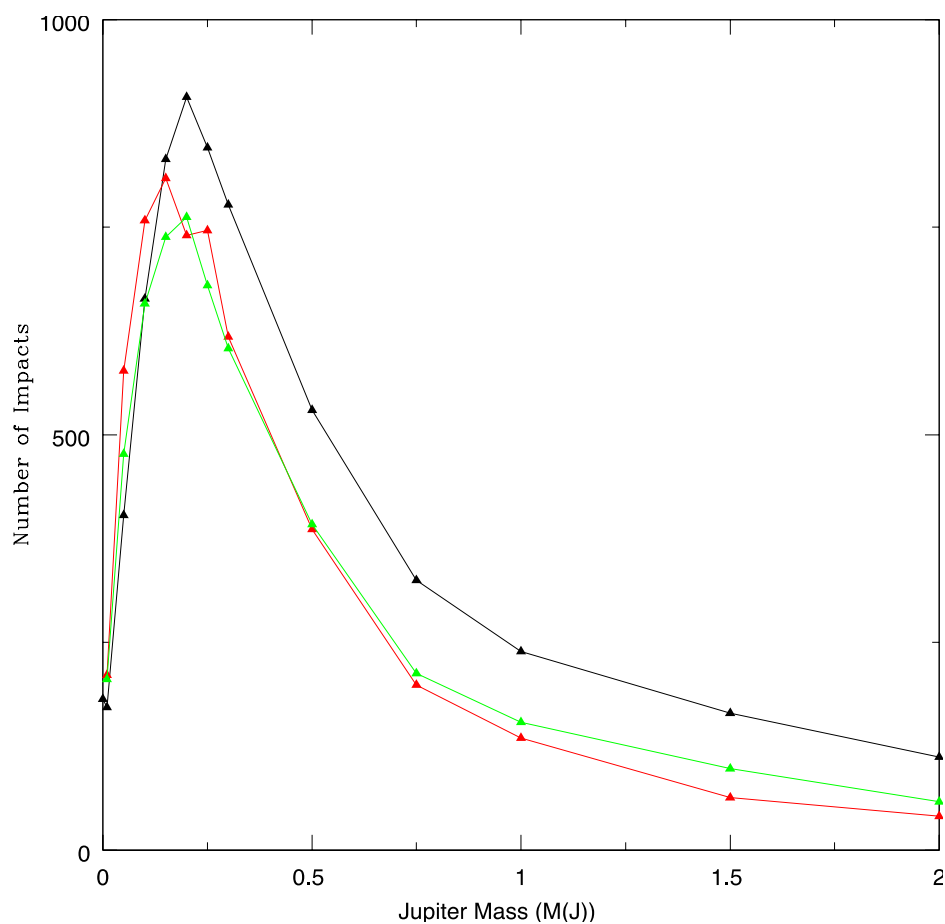

**Figure 5.** The impact flux at the Earth as a function of Jupiter mass for the short-period comets. As was the case for the near-Earth asteroids (Figure 3), the data plotted in black show the results from our earlier work, Horner & Jones, 2009, with the giant planet moving on an orbit that was barely inclined to the plane of the ecliptic (1.31°). The data plotted in green show the results for orbits of moderate (5°) inclination, while those in red show the scenario in which the giant planet was placed on a highly inclined (25°) orbit. In stark contrast to the results shown in Figure 3, it turns out that a more inclined giant planetary orbit results in a reduced impact rate at the Earth for the short-period comets.

Interestingly, and in stark contrast to the results described above, for the effect of orbital inclination on the impact flux from the asteroid belt, increasing "Jupiter"'s orbital inclination has only a small effect on the impact flux experienced from the short-period comets. Figure 5 shows that, above a "Jupiter" mass of ~0.1 $M_J$, the impact flux from those objects is highest for the scenario in which "Jupiter" has the lowest orbital inclination. The impact fluxes for the moderate and highly inclined scenarios are broadly similar, with impact rates slightly lower than those experienced in the low inclination scenario. Whilst there is therefore evidence of some modulation of the impact rate as a result of "Jupiter"'s orbital inclination, it is nowhere near as dramatic as the variation observed for objects sourced from the asteroid belt.

**Discussion and future work**

It is interesting to note that all of our results (Figures 1 – 5) follow the same general pattern to that which was presented in our earlier investigations into the influence of giant planet mass on the impact rate from the short-period comets and the near-Earth asteroids (Horner & Jones, 2009, 2008b). Regardless of the orbital eccentricity and inclination of the giant planet (with the proviso that the scenarios we tested were limited by the constraint of not pushing the planetary system into an unstable state), the impact rate experienced by the Earth is highest for a giant planet of mass ~0.25 $M_J$, for both test populations. This seems to be a recurring theme – while Jupiter itself is most likely neither a Friend nor Foe, it seems that were it swapped for a Saturn-mass planet, the Earth would experience a significantly enhanced flux of impacts over that we see today.

Overall, it is apparent that an increased giant planet orbital eccentricity leads to an increased impact flux upon the Earth for both the near-Earth asteroids and the short-period comets. This result is not unexpected, since an increased orbital eccentricity will lead to Jupiter's gravitational reach extending further from its orbit in both the Sunward and anti-Sunward directions, effectively giving it a greater zone of gravitational dominance[3]. This will, in turn, make it easier for Jupiter to nudge a greater flux of objects onto Earth-crossing orbits, thereby increasing the impact flux. Interestingly, when we consider scenarios in which our Jupiter is placed on an orbit of significantly greater eccentricity (0.1), a secondary peak in the impact flux becomes apparent for both the near-Earth asteroids and the short-period comets – placing our Jupiter at a local minimum in the impact flux distribution for such orbits. This is, in part, due to the combination of the increased gravitational "reach" of the planet, as described above. Increasing the eccentricity of the planet's orbit is not the only way to increase its reach, however. It is readily apparent that, as the mass of the planet increases, so too does the distance at which it can strongly perturb passing objects. This effect has been discussed in some depth in our earlier work. When this is taken into account, the overall effect on the planet's gravitational reach resulting from its increased orbital eccentricity is clearly greater for scenarios involving a more massive "Jupiter". This effect is particularly important for the case of the injection of material from the asteroid belt. At the highest Jupiter masses, the planet begins to whittle away at the outer-edge of the belt in a way that is not observed when the planet is less massive, enhancing the impact flux by more than might otherwise have been expected. In addition, though, a secondary (and more important) effect plays a role for the injection of material from the asteroid belt. Increasing the orbital eccentricity of "Jupiter"'s orbit causes the location of the $\nu_6$ secular resonance to shift to larger heliocentric distances, and also causes some broadening of the region affected by the resonance. This effect can be clearly seen in Figure 2, and in all of the subsidiary figures presented in Appendix I, except for that showing the 0.01 $M_J$ case. In the case of the scenario featuring a "Jupiter" of mass 1.5 $M_J$ (Figure 2), the $\nu_6$ secular resonance has all but left the inner edge of the asteroid belt for the case where that planet's orbital eccentricity is low. Increasing the eccentricity causes the resonance to instead be located within the inner reaches of the belt, allowing it to significantly destabilise a region that would otherwise have been relatively untouched. This, in turn, results in a significant increase in the observed impact flux at Earth, causing the observed secondary peak at 1.5 $M_J$.

---

[3] To illustrate this enhanced reach, let us consider two "Jupiter"s, one moving on a perfectly circular orbit at $a$=5.2 AU, the other moving on an orbit with eccentricity 0.1, and the same semi-major axis. As a result of its orbital eccentricity, that planet will pass through perihelion at 4.68 AU, and will reach aphelion at 5.72 AU. If we assume, for the sake of argument, that this particular "Jupiter"'s gravitational reach extends 1 AU from the planet, then the planet moving on a circular orbit can strongly perturb objects between 4.2 and 6.2 AU from the Sun. By contrast, the planet moving on an eccentric orbit can perturb objects at 3.68 AU from the Sun, at perihelion, and out to 6.72 AU at aphelion, a significantly enhanced reach.

The results of our investigation of planetary systems with a moderately or highly inclined giant planet are somewhat more surprising. The most straightforward results are those for the short-period comets, where the maximum impact flux is highest for the scenarios in which Jupiter moves in its present inclination, $i_J$ = 1.31°, close to the invariant plane of the planetary system. At higher orbital inclinations, the impact flux is noticeably lower, aside from at very low Jupiter masses, and high Jovian inclinations, where the flux appears to be slightly elevated over that for the co-planar case (Figure 5). It also is interesting to note in passing that the most inclined scenario tested, with $i_t$ = 25°, shows hints of a double-peaked impact distribution reminiscent of the impact flux observed for the Asteroids.

In the case of the influence of planetary inclination on the impact flux resulting from the asteroids, we see something dramatic (Figure 4). For a moderate inclination (5°), the impact flux at Earth is greatly enhanced over that experienced in those runs where the massive planet moves close to the plane of the ecliptic. Far more dramatic, however, is the result for the highly inclined scenarios ($i_J$ = 25°). Here, the impact flux is essentially almost saturated, close to the 100% mark, for planets of mass between ~0.2 and 0.5 $M_J$. We remind the reader, at this point, that the Earth used in our integrations was inflated to a radius of one million kilometres, in order to boost the impact rate recorded. In this case, it seems that such drastic measures were certainly not needed! Rather than inferring that a highly inclined giant will lead to an impact rate from asteroids so great that life could not develop on a planet in such a system, this impact rate is sufficiently high that it instead reveals that, in such systems, no stable asteroid belt could exist. The region occupied by the asteroids in our runs is swept clean incredibly efficiently. Whilst this would mean that the impact rate on objects in the inner reaches of such a planetary system would be punishingly high during the latter stages of planetary formation, it seems reasonable to assume that, by the time the system settled down, such systems would simply not contain a reservoir of objects analogous to the asteroid belt, and so the single largest source of hazardous objects to our Earth would not exist. In retrospect, this result is perhaps not that surprising. When one examines plots of the distribution of the Asteroids within our own Solar System (such as those presented in Figure 1 of O'Brien & Sykes, 2011), it is immediately apparent that there are far broader regions of dynamical stability that can be occupied by asteroids at low orbital inclinations than at high inclinations. If we purely consider a three-body system involving the Sun, Jupiter, and an asteroid, then it is clear that an asteroid moving with an orbit inclined by 25° in a system where Jupiter moves in the plane of the ecliptic is essentially the same as a scenario where the asteroid is moving in the plane of the ecliptic, and Jupiter is inclined by 25°. Whilst those plots are clearly not truly representative of the results of our work, they indicate that the fraction of the semi-major axis space occupied by the asteroid belt that is dynamically stable falls dramatically as the inclination of an asteroid's orbit, with respect to that of Jupiter, is increased. We note, in passing, that the $\nu_6$ secular resonance can again clearly be seen in such plots, and constitutes a significant contribution to the destabilisation of the belt at increased inclinations.

**Conclusions**

We have revisited our earlier studies of the influence of Jupiter on the impact flux experienced by the Earth for the two main populations of potentially hazardous objects – the asteroids and the short-period comets. In this work, we have examined the role played by the orbital inclination and eccentricity of Jupiter-like planets. In the case of orbital eccentricity, we find that giant planets on more eccentric orbits would lead to increased impact fluxes from both cometary and asteroidal bodies, when compared to planets on less eccentric orbits. By contrast, increasing the orbital inclination of Jupiter-like planets actually acts to reduce the impact flux from short-period comets that would be experienced by potentially habitable planets in the same system. Increasing the Jovian orbital inclination has a dramatic impact on the asteroid belt, with moderately inclined "Jupiter"s leading to a greatly increased impact flux on the terrestrial worlds when compared to those moving in the plane of the planetary system. When the giant planet is placed on a highly inclined orbit (25°), however, the asteroid belt is so destabilised that the entire belt would be removed on astronomically short time scales. In such systems, therefore, the eventual impact flux on the terrestrial planets would be far lower than that seen in our own Solar System, since no reservoir would exist to source fresh asteroids to the inner Solar System. Our results once again highlight the danger of simply assuming that a giant planet will act as a shield to potentially habitable worlds in its system. This reinforces the conclusions of Horner & Jones, 2010a that, in the future, any planetary systems found to host potentially habitable exo-Earths should be examined individually in great detail to ascertain the various factors that could impact upon their habitability, before any conclusions are drawn about that planet's suitability as a target in the search for life beyond our Solar System.

## Acknowledgements

JH gratefully acknowledges the financial support of the Australian government through ARC Grant DP0774000. The authors would also like to thank an anonymous referee for a number of helpful comments and suggestions, which improved the overall presentation and flow of this work

**Appendix I:** Final Asteroid Belts for low (0.01, plotted in green), medium (0.0488, plotted in black) and high (0.10, plotted in red) eccentricities. We remind the reader that each distribution begins at zero on the y-axis, and that the medium and high eccentricity cases have simply been displaced vertically, to facilitate the direct comparison between the three cases shown.

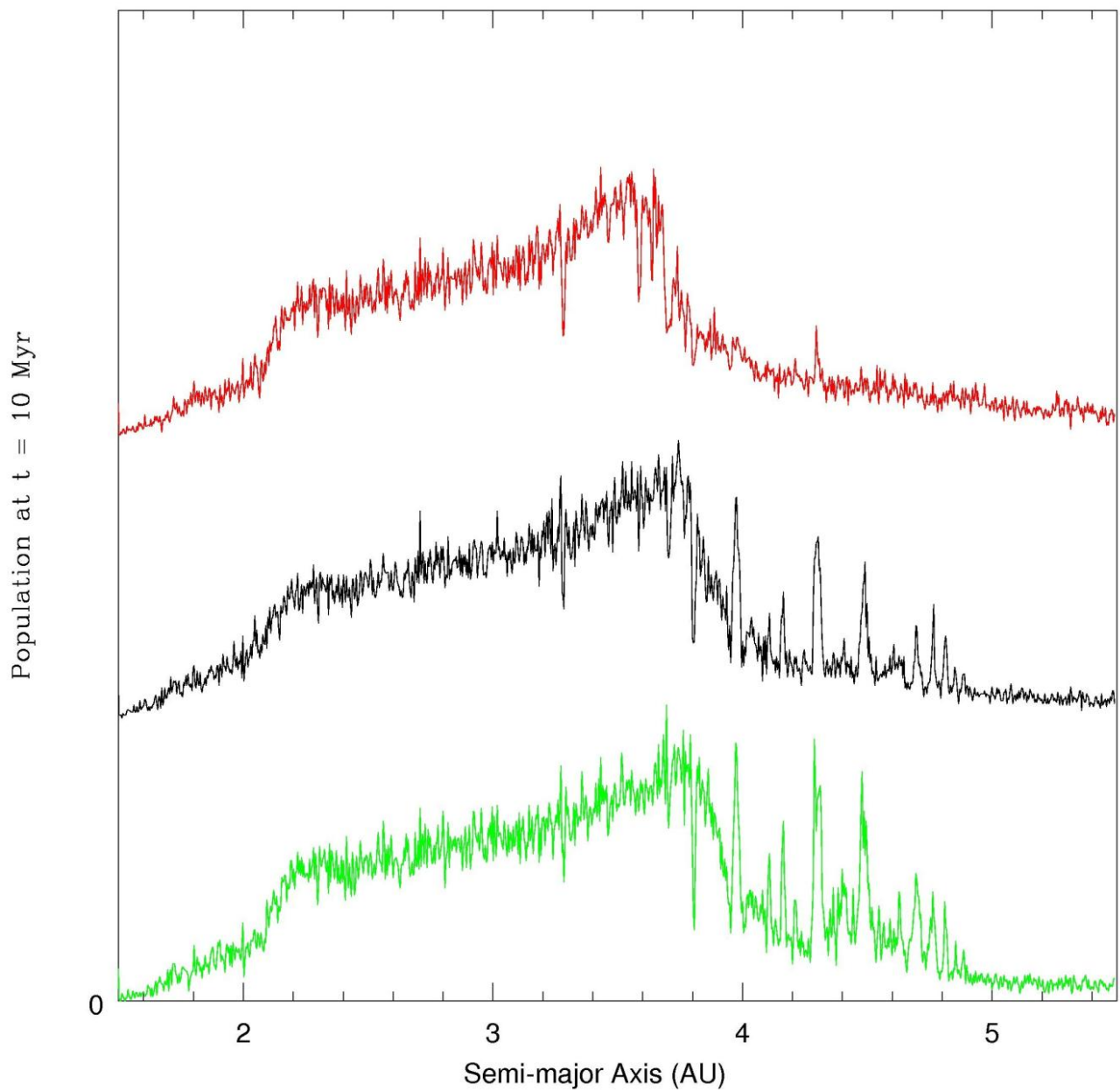

$M = 0.01\ M_J$

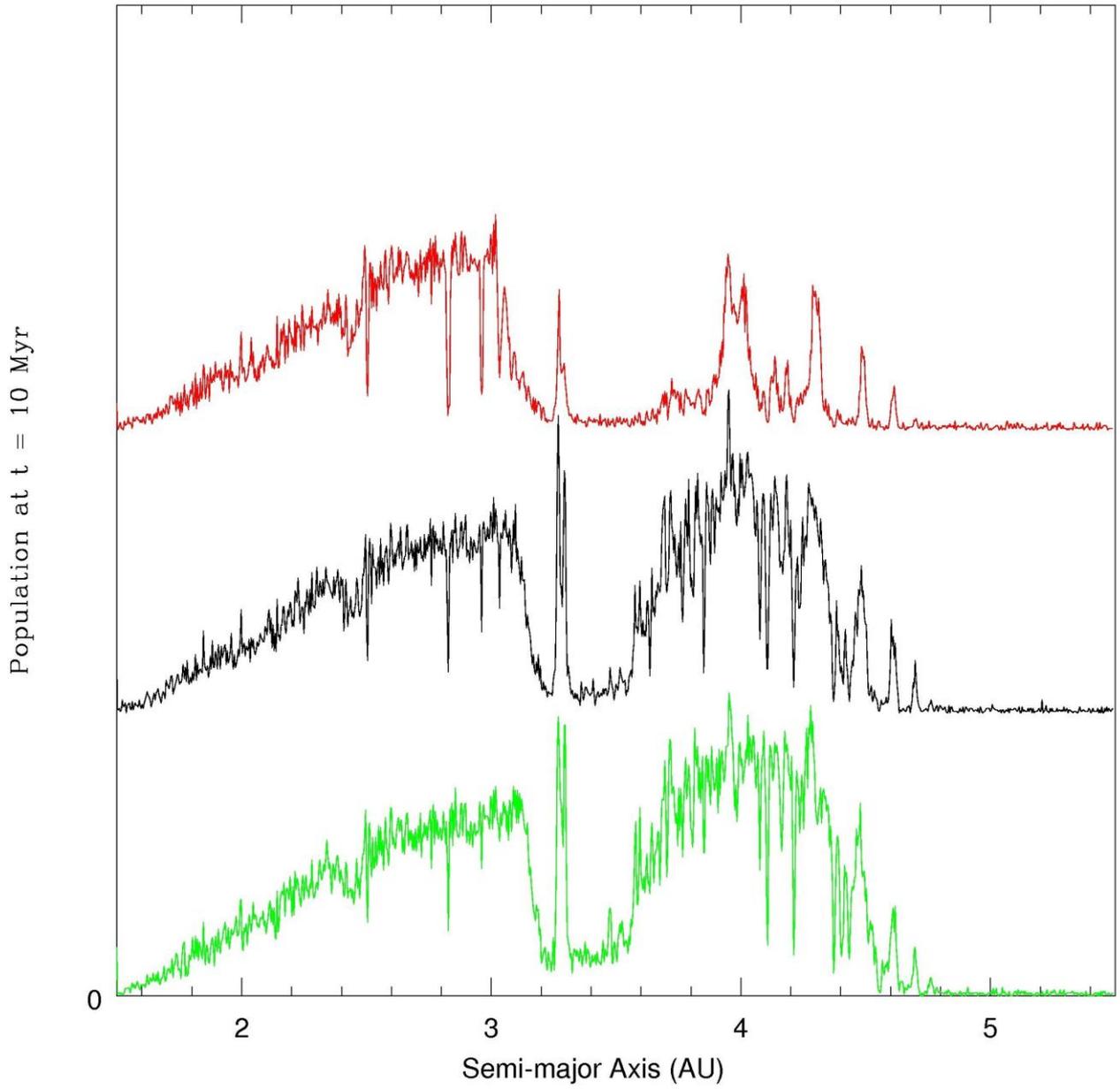

$M = 0.05\ M_J$

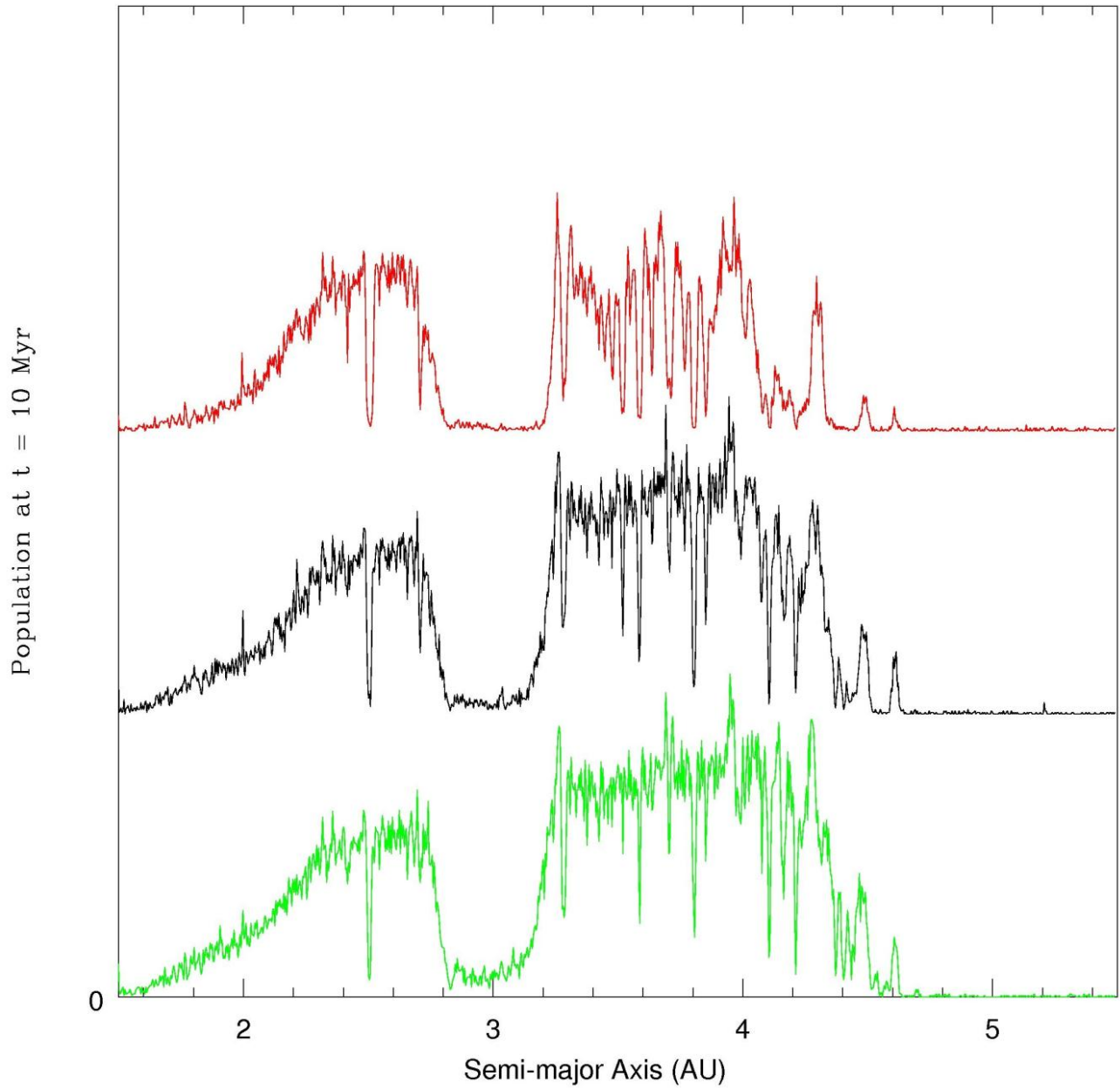

$M = 0.10\ M_J$

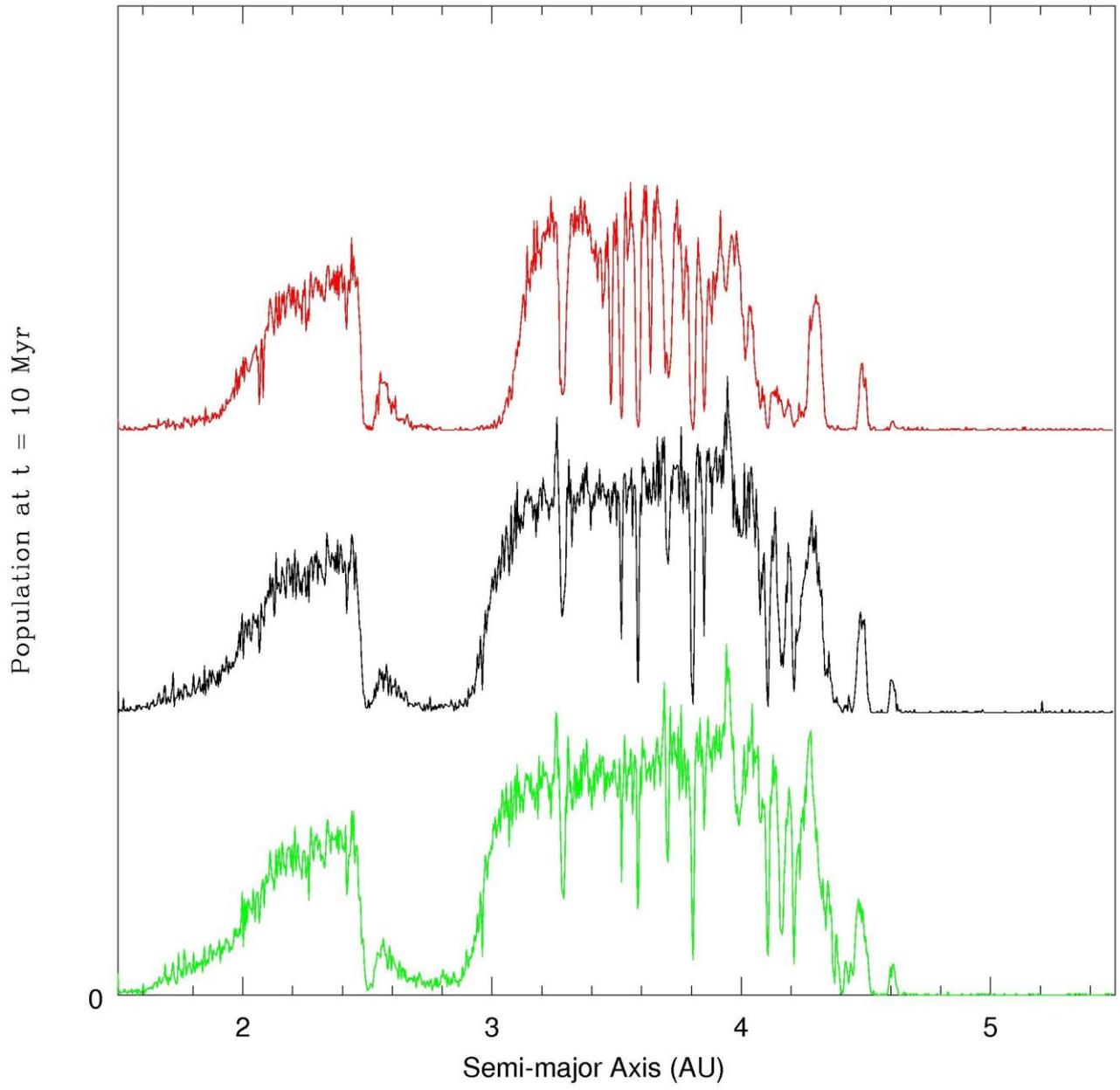

$M = 0.15\ M_{\rm J}$

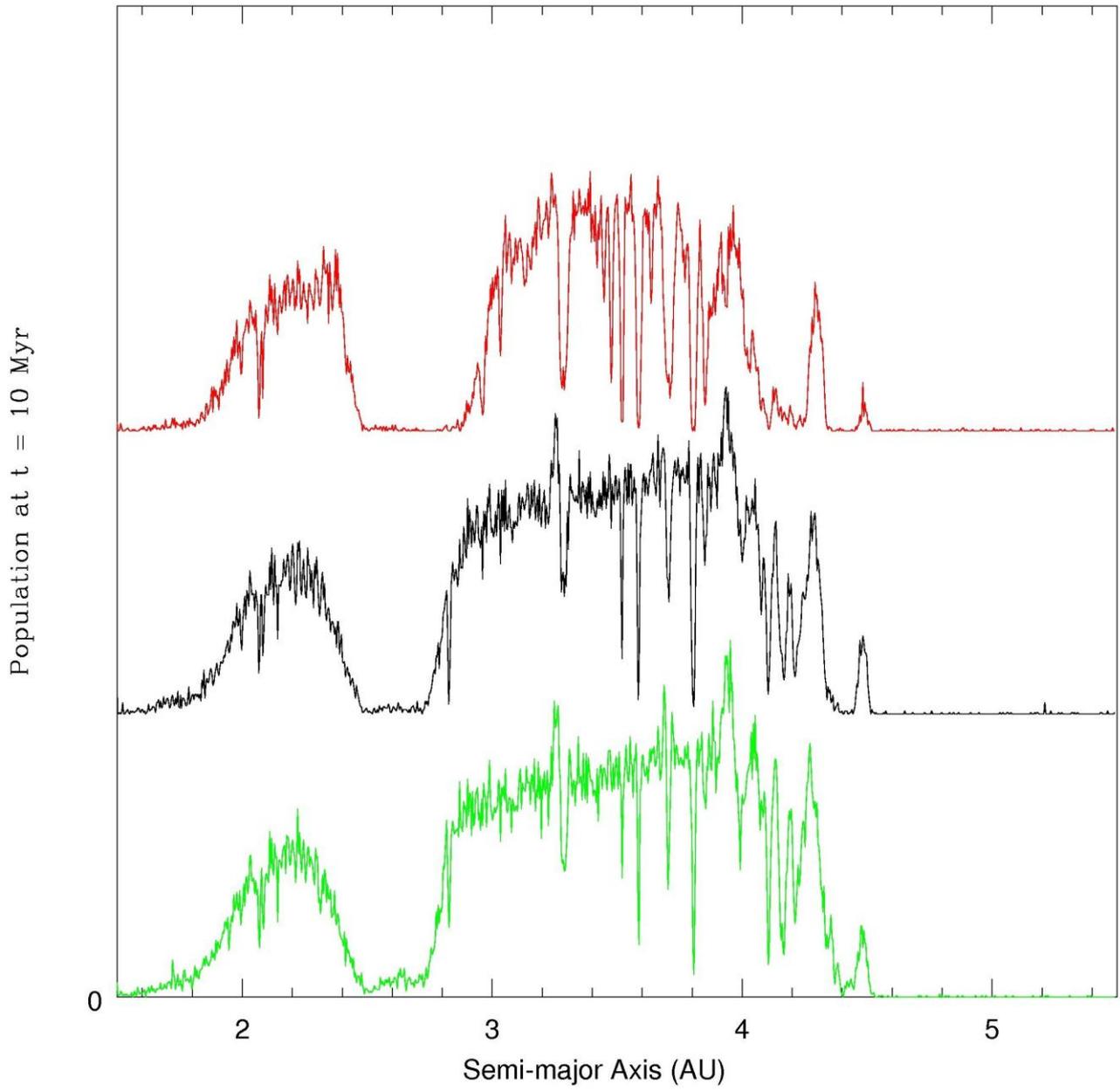

$M = 0.20\ M_J$

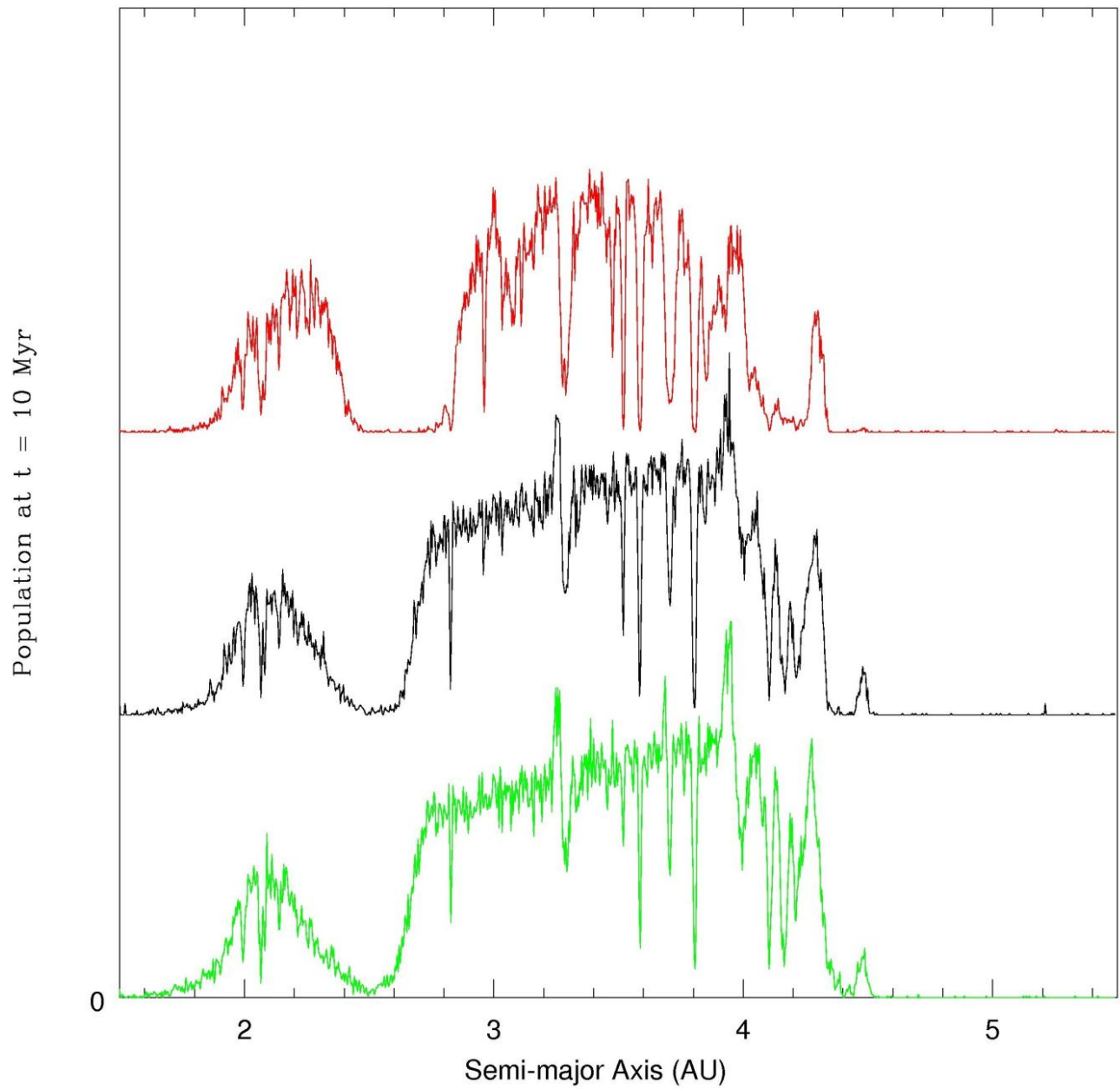

$M = 0.25\ M_J$

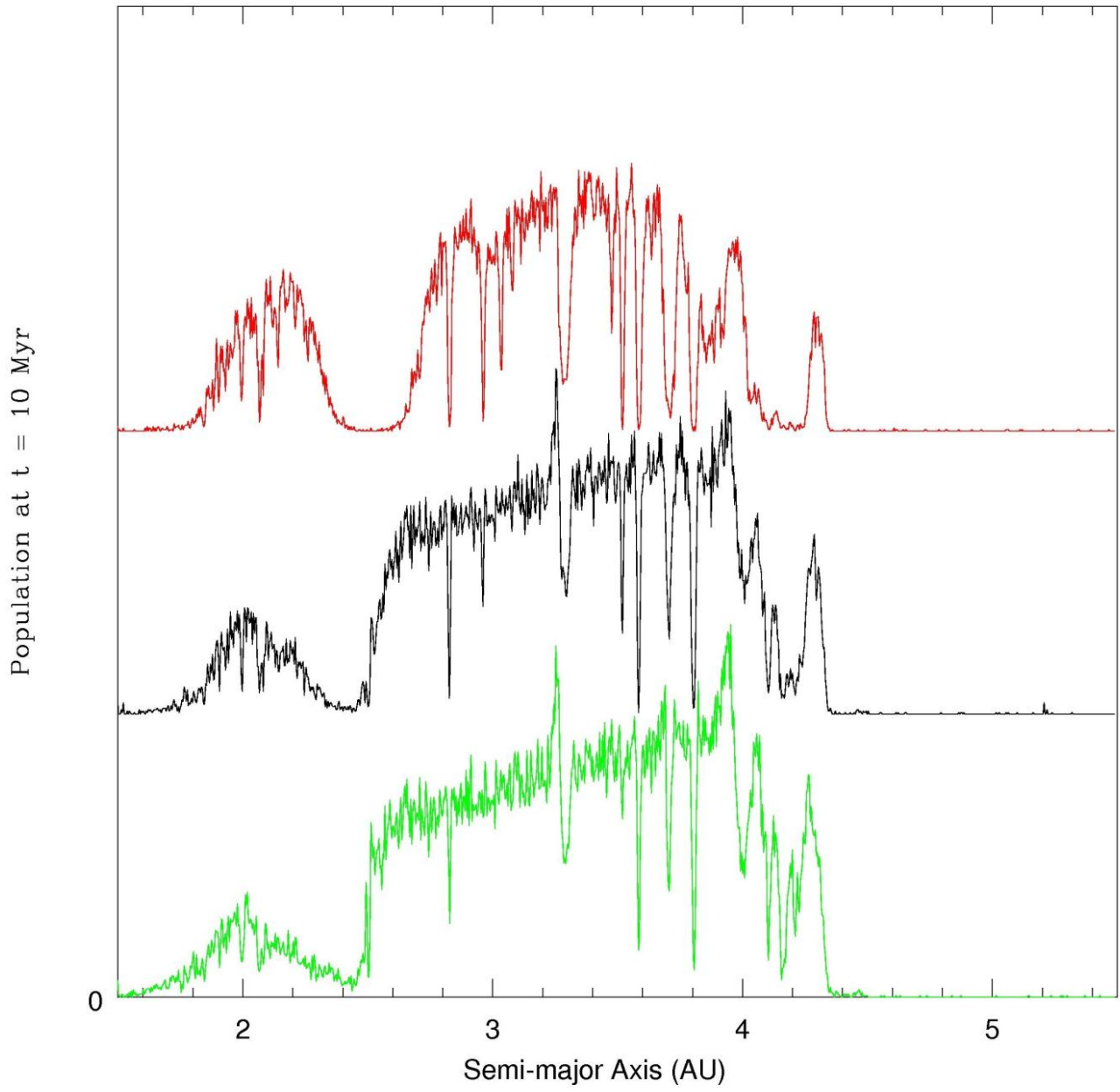

$M = 0.33\ M_J$

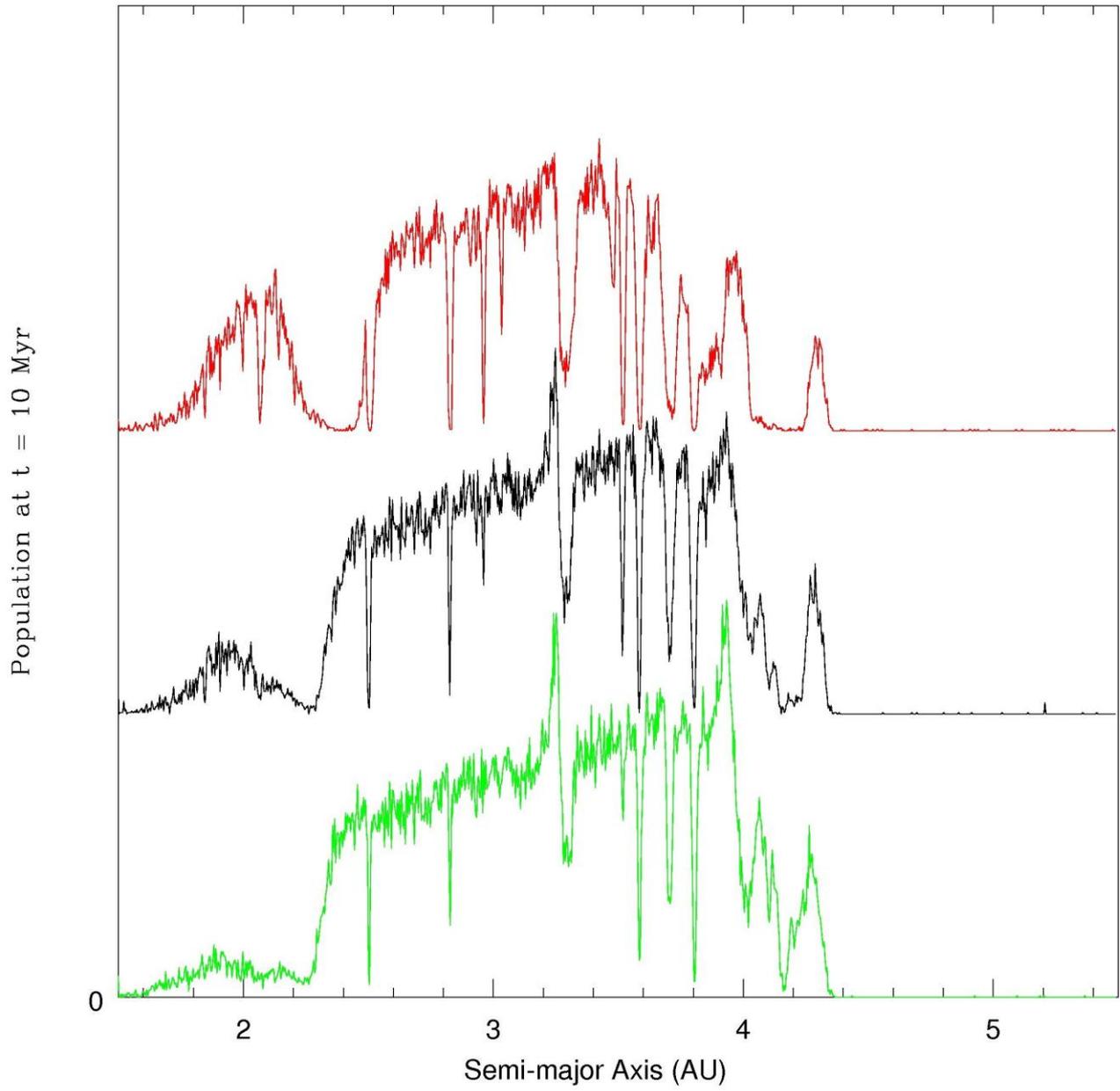

$M = 0.50\ M_J$

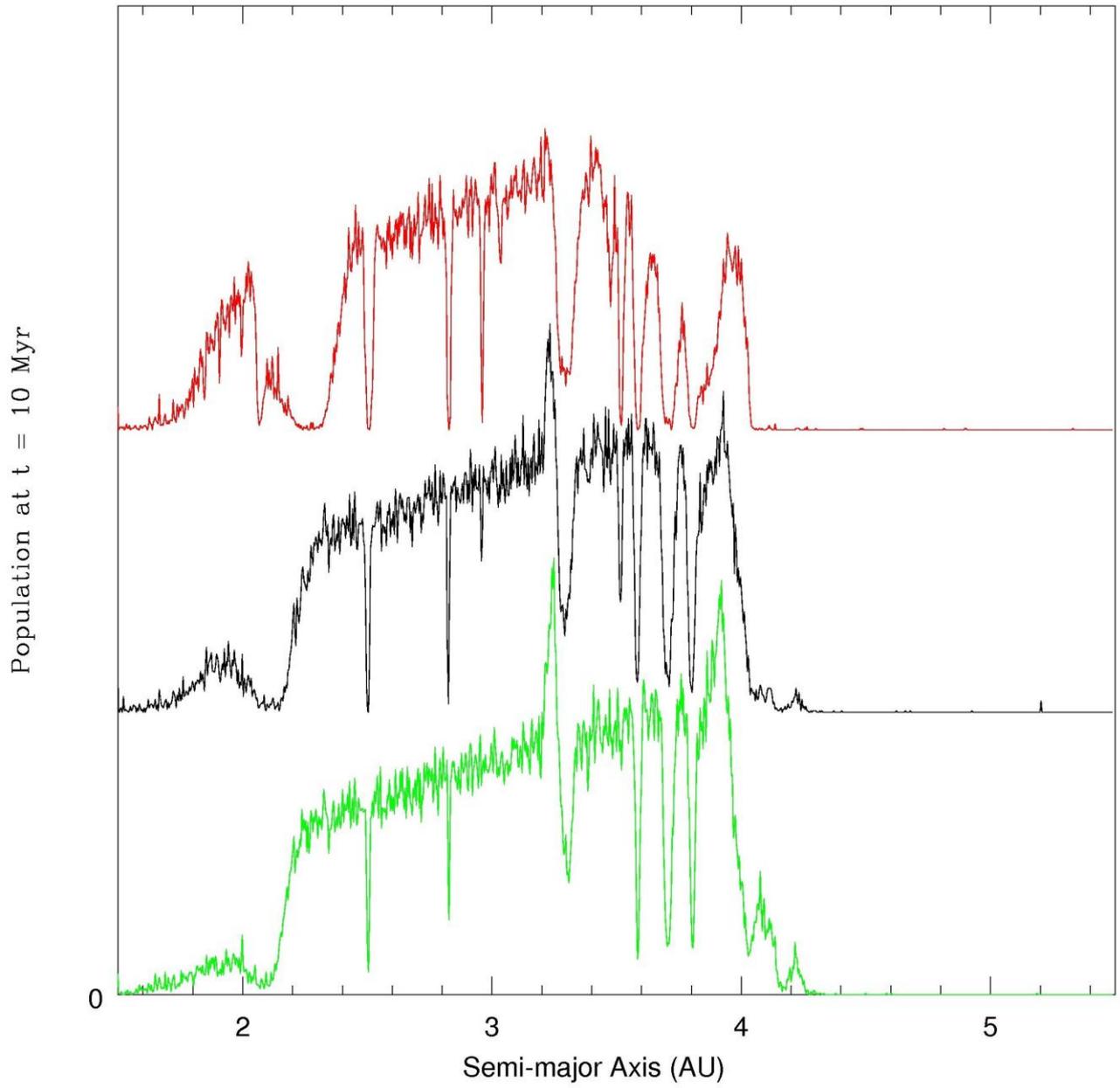

$M = 0.75\ M_J$

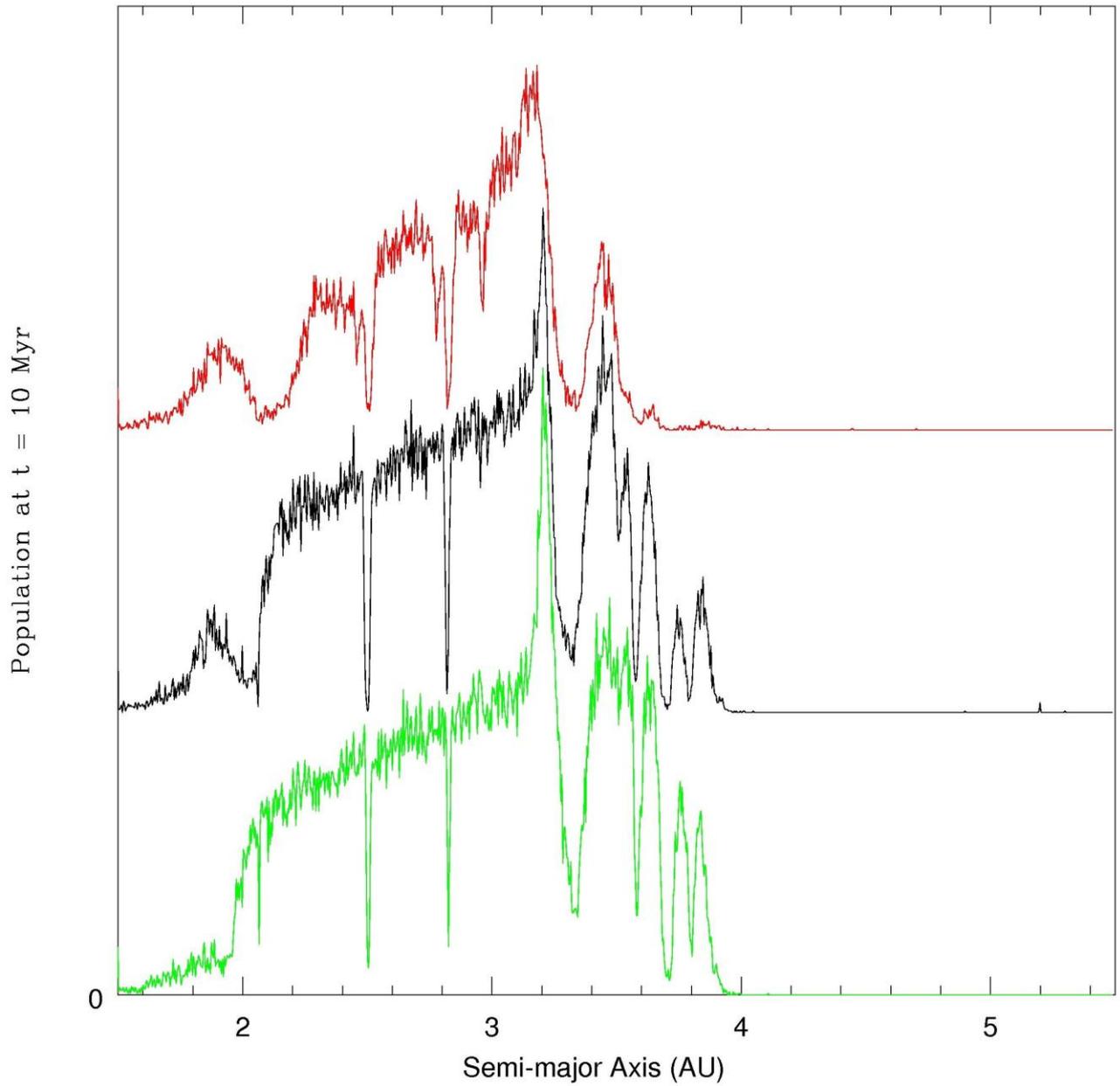

$M = 1.00\ M_J$

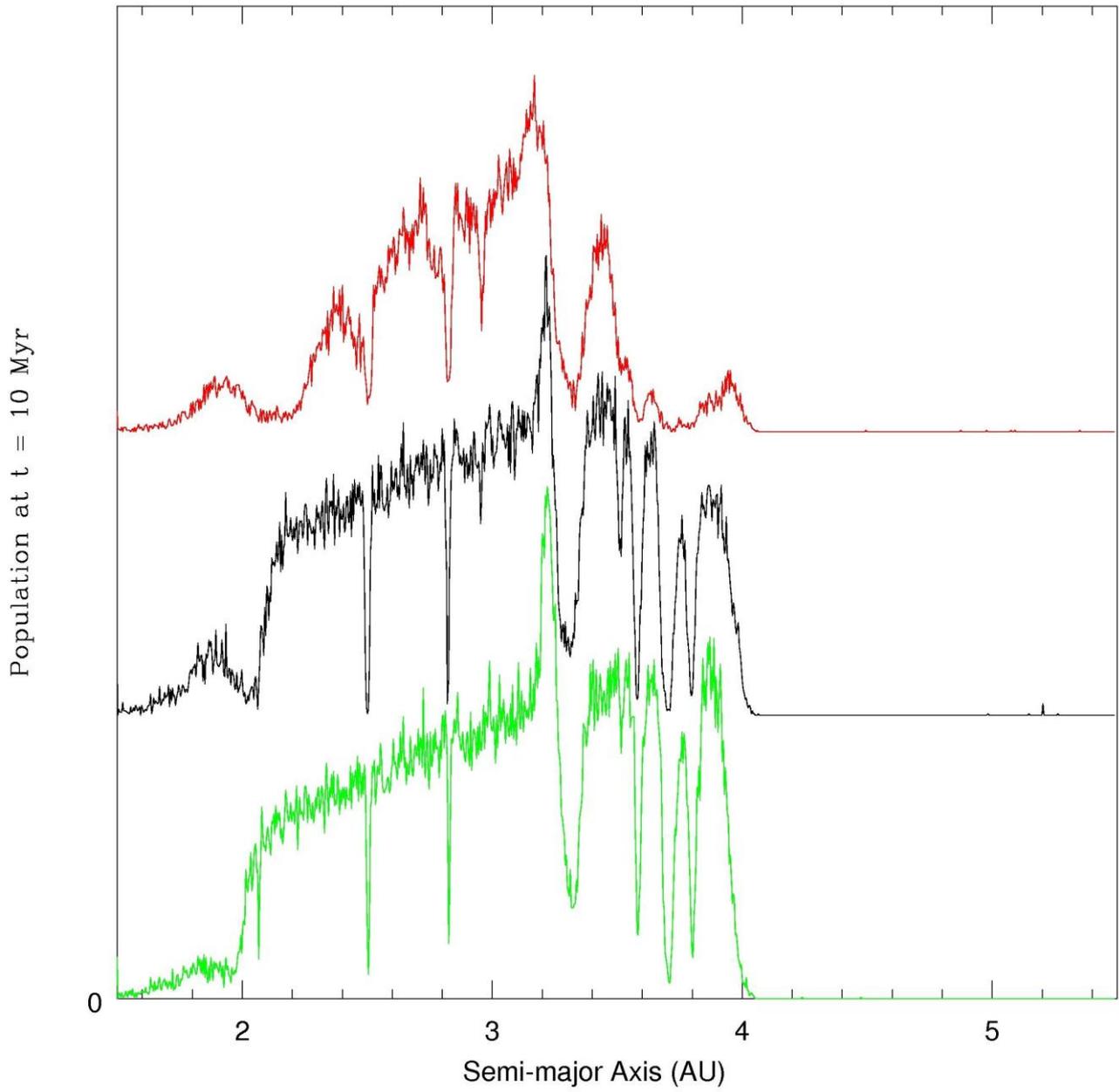

$M = 1.50\ M_J$

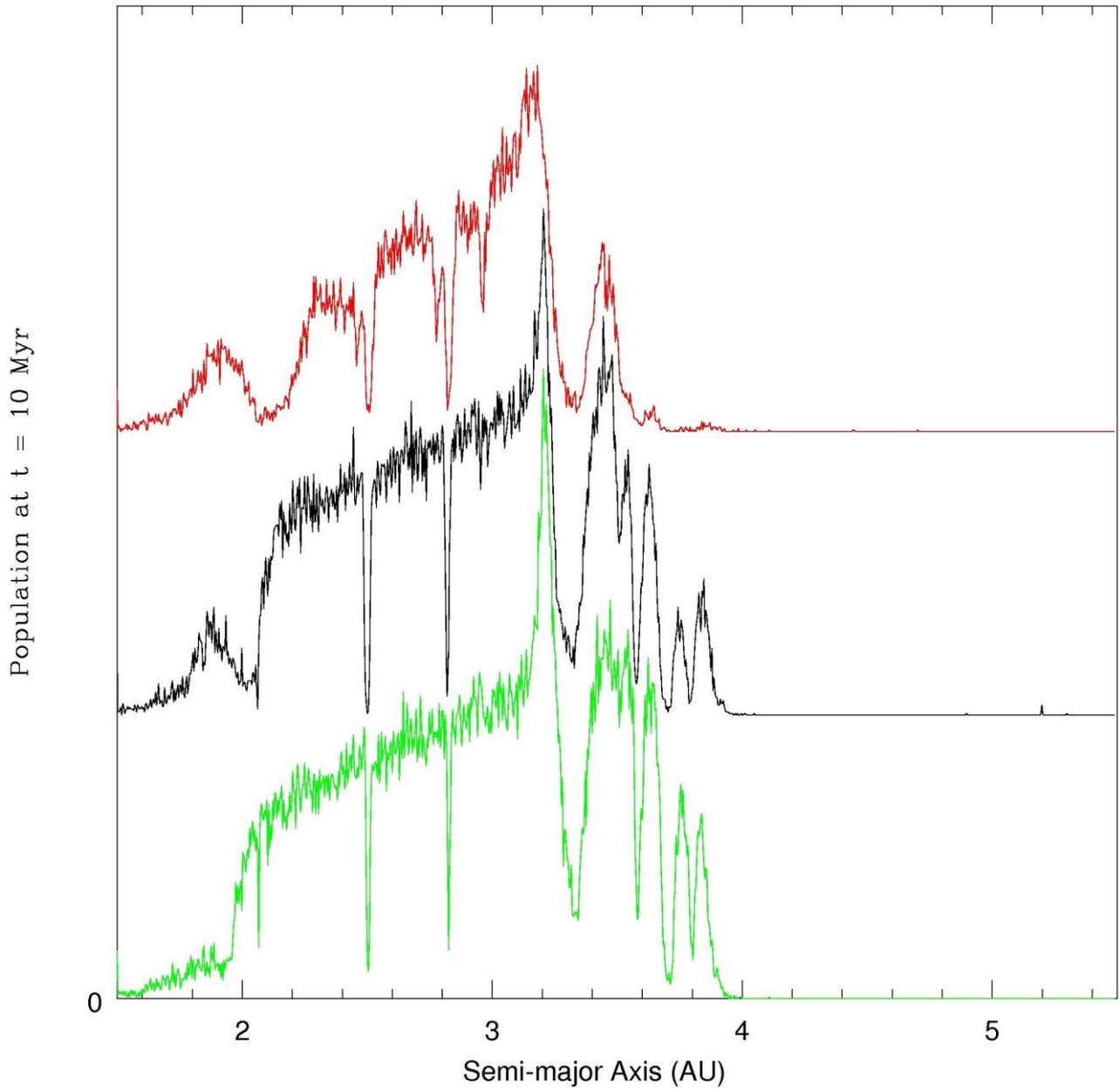

$M = 2.00 \, M_J$